\documentclass[twocolumn]{aastex62} 
\usepackage{savesym}
\savesymbol{tablenum}
\usepackage{siunitx}
\restoresymbol{SIX}{tablenum}
\usepackage{url}
\usepackage{comment}
\usepackage[english]{babel}
\usepackage{xspace}
\usepackage[xindy, toc, hyperfirst=false, nolist, nostyles, sanitize={name=false,description=false,symbol=false}]{glossaries}
\glsdisablehyper

\newglossaryentry{vrad}{name={radial velocity~}, text={radial velocity}, symbol={\ensuremath{v_\textrm{rad}}}, description={radial velocity}, sort=vrad}
\newglossaryentry{vrot}{name={stellar rotation~}, name={stellar rotation}, symbol={\ensuremath{v_\textrm{rot}}}, description={radial velocity}, sort=vrot}

\newcommand{\xray}{X-ray}

\newglossaryentry{angstrom}{name=\AA, description={unit of length $10^{-10}$\,m}, sort=angstrom}
\newglossaryentry{nir}{name=NIR,description={near infrared},first = {near infrared (NIR)}}
\newglossaryentry{psf}{name=PSF,description={point-spread function},first = {point-spread function (PSF)}}
\newglossaryentry{fwhm}{name=FWHM,description={Full Width Half Maximum},first = {FWHM}}
\newglossaryentry{rms}{name=RMS,description={Root Mean Square},first = {RMS}}
\newglossaryentry{signalnoise}{name=S/N,description={signal to noise}}
\newglossaryentry{uv}{name=UV,description={ultra violet},first = {ultra violet (UV)}}
\newglossaryentry{halpha}{name=\ensuremath{\textrm{H}\alpha}, description={First line of the Balmer series at 6563\,\AA}, sort=halpha}
\newglossaryentry{mgb}{name={Mg \textsc{i} b}, description={Triplet at 5167\,\AA, 5173\,\AA and 5184\,\AA}}
\newglossaryentry{sobolevapprox}{name={Sobolev approximation}, description={Lines are approximation with an infinitley thin interaction region \citep[e.g. no broadening][]{1960mes..book.....S}}, first={Sobolev approximation }}
\newglossaryentry{radeq}{name={radiative equilibrium}, description={The net flux of energy between matter and radiation field is zero}}
\newglossaryentry{nebularapprox}{name={nebular approximation}, description={Assumes that the plasma condition are controlled by a central radiation source. The radiation field decreases with the distance to the source by geometrical dilution. See \citet{1978stat.book.....M} for details}}
\newglossaryentry{modnebularapprox}{name={modified nebular approximation}, description={In contrast to \gls{nebularapprox} where only geometrical dilution is taken into account, the modified nebular approximation also takes dilution by other radiative processes into account }, first={modified nebular approximation}, parent=nebularapprox}
\newglossaryentry{thompsonscat}{name={Thomson scattering}, description={Scattering of photons on low energy electrons}}
\newglossaryentry{lte}{name={LTE}, description={Local Thermodynamic Equilibrium}, first={local thermodynamic equilibrium (LTE)}}
\newglossaryentry{lsr}{name={LSR}, description={Local Standard of Rest}, first={\textit{local standard of rest} (LSR)}}
\newglossaryentry{mc}{name={MC}, description={Monte Carlo}, first={\textit{Monte Carlo} (MC)}}
\newglossaryentry{wcs}{name={WCS}, description={world coordinate system}, first={world coordinate system (WCS)}}
\newglossaryentry{cmf}{name=CMF, text=CMF, first=Comoving Frame (CMF henceforth), description={Comoving Frame}}

\newglossaryentry{uvoir}{name=UVOIR, text=UVOIR, first=UV/optical/Near-IR (UVOIR), description={UV/optical/Near-IR}}

\newglossaryentry{sfit}{name=SFIT, text=\textsc{sfit}, description={spectral fitting program for hot stars \citep{2001A&A...376..497J}}, first={\textsc{sfit} \citep{2001A&A...376..497J}}}
\newglossaryentry{iraf}{name=IRAF, text=\textsc{iraf}, description={Image Reduction and Analysis Facility maintained by NOAO}, first={\textsc{iraf}\protect\footnote{IRAF: the Image Reduction and Analysis Facility is distributed by the National Optical Astronomy Observatory, which is operated by the Association of Universities for Research in Astronomy (AURA) under cooperative agreement with the National Science Foundation (NSF).}}}
\newglossaryentry{pyraf}{name=PyRAF, text=\textsc{PyRAF}, description={Python wrap of \gls{iraf} maintained by STSCI}, first=\textsc{PyRAF} \protect\footnote{PyRAF is a product of the Space Telescope Science Institute, which is operated by AURA for NASA.}}
\newglossaryentry{astropy}{name=ASTROPY, text=\textsc{astropy}, description=\textsc{astropy} framework, first = \textsc{astropy} \citep{2013A&A...558A..33A}}
\newglossaryentry{numpy}{name=NUMPY, text=\textsc{numpy}, description=\textsc{numpy} framework, first = \textsc{numpy} \citep{walt2011numpy}}
\newglossaryentry{scipy}{name=SCIPY, text=\textsc{scipy}, description=\textsc{scipy} framework, first = \textsc{scipy} \citep{Jones:2001fk}}
\newglossaryentry{matplotlib}{name=matplotlib, text=\textsc{matplotlib}, description=\textsc{matplotlib} framework, first = \textsc{matplotlib} \citep{hunter2007matplotlib}}
\newglossaryentry{pandas}{name=pandas, text=\textsc{pandas}, description=\textsc{pandas} framework, first = \textsc{pandas} \citep{mckinney2010data}}
\newglossaryentry{ipython}{name=ipython, text=\textsc{ipython}, description=\textsc{ipython} framework, first = \textsc{ipython} \citep{perez2007ipython}}
\newglossaryentry{jupyter}{name=jupyter, text=\textsc{jupyter}, description=\textsc{jupyter} framework, first = \textsc{jupyter} \citep{kluyver2016jupyter,perez2015project,ragan2014jupyter}}
\newglossaryentry{aplpy}{name=aplpy, text=\textsc{aplpy}, description=\textsc{aplpy} framework, first = \textsc{aplpy} \citep{2012ascl.soft08017R}}
\newglossaryentry{nltk}{name=nltk, text=\textsc{nltk}, description=\textsc{nltk} framework, first = Natural Language ToolKit \citep[\textsc{NLTK};][]{bird2009natural}}
\newglossaryentry{scikit-learn}{name=scikit-learn, text=\textsc{scikit-learn}, description=\textsc{scikit-learn} framework, first = \textsc{scikit-learn} \citep[][]{scikit-learn}}
\newglossaryentry{moog}{name=MOOG,text={\textsc{moog}}, description={spectral synthesis software \citep{1973ApJ...184..839S}}, first={\textsc{Moog} \citep{1973ApJ...184..839S}}}
\newglossaryentry{atlas9}{name=ATLAS9,description={grid of stellar atmospheres \citep{2004astro.ph..5087C}}, first={ATLAS9 \citep{2004astro.ph..5087C}}}
\newglossaryentry{vald}{name=VALD,description={Vienna Atomic Line Database \citep{2000BaltA...9..590K}}, first={Vienna Atomic Line Database \citep[VALD;][]{2000BaltA...9..590K}}}
\newglossaryentry{sextractor}{name=SExtractor, text=\textsc{SExtractor}, description={Source Extractor photometry program \citep{1996A&AS..117..393B}}, first={\textsc{SExtractor} \citep{1996A&AS..117..393B}}}

\newglossaryentry{swarp}{name=SWarp, text=\textsc{SWarp}, description={SWarp \citep{2002ASPC..281..228B}}, first={\textsc{SWarp} \citep{2002ASPC..281..228B}}}

\newglossaryentry{astrodrizzle}{name=AstroDrizzle, text=\textsc{AstroDrizzle}, description={AstroDrizzle \citep{2012drzp.book.....G}}, first={\textsc{AstroDrizzle} \citep{2012drzp.book.....G}}}

\newglossaryentry{idl}{name=IDL,text={\textsc{idl}}, description={Interactive Data Language}}
\newglossaryentry{makee}{name=MAKEE,text=\textsc{makee}, description={MAuna Kea Echelle Extraction by Tom Barlow available}}
\newglossaryentry{minuit}{name=MINUIT,text={\textsc{minuit}}, description={collection of numerical optimization tools \citep{James:1975dr}}}
\newglossaryentry{migrad}{name=MIGRAD,text={\textsc{migrad}}, description={numerical gradient optimization tools - part of \gls{minuit}}}
\newglossaryentry{dolphot}{name=DOLPHOT, text=\textsc{dolphot}, description=photometry package for HST, first=\textsc{dolphot} \citep{2000PASP..112.1383D}}
\newglossaryentry{synphot}{name=synphot, text={\textsc{synphot}}, description={synthetic photometry package from STSCI}, first={\textsc{synphot}\protect\footnote{\textsc{synphot} is a product of the Space Telescope Science Institute, which is operated by AURA for NASA.}}}
\newglossaryentry{chianti}{name=CHIANTI, text=CHIANTI, description= CHIANTI Database 7.1, first =CHIANTI 7.1 \citep{1997A&AS..125..149D,2012ApJ...744...99L}}
\newglossaryentry{synpp}{name=SYNPP, text=SYN++, description= SYN++ software, first =SYN++ \citep{2011PASP..123..237T}}
\newglossaryentry{tardis}{name=TARDIS, text=\textsc{tardis}, description= TARDIS MC code, first = {\textsc{tardis} \citep{2014MNRAS.440..387K}}}

\newglossaryentry{artis}{name=ARTIS, text=\textsc{artis}, description= ARTIS MC code, first = \textsc{artis} \citep{2009MNRAS.398.1809K}}
\newglossaryentry{sedona}{name=SEDONA, text=\textsc{sedona}, description= Sedona MC code, first = \textsc{sedona} \citep{2006ApJ...651..366K}}
\newglossaryentry{mlmc}{name=MLMC, text=ML93, description= Mazzali Lucy Monte Carlo, first ={Mazzali \& Lucy (1993, ML93) code}}
\newglossaryentry{starkit}{name=STARKIT, text=\textsc{starkit}, description= TARDIS MC code, first = {\textsc{starkit} \citep{wolfgang_kerzendorf_2015_28016}}}

\newglossaryentry{pyne}{name=PYNE, text=\textsc{pyne}, description= PYNE code, first = {\textsc{pyne} \citep{Scopatz2012a}}}
\newglossaryentry{multinest}{name=MULTINEST, text=\textsc{MultiNest}, description=MultiNest, first={\textsc{MultiNest} \citep{2009MNRAS.398.1601F}}}

\newglossaryentry{ads}{name=ADS ,description=ADS, first={NASA Astrophysics Data System (ADS) \citep{2000A&AS..143...41K}}}

\newglossaryentry{2mass}{name=2MASS,description={Two Micron All Sky Survey \citep{2006AJ....131.1163S}}, first={Two Micron All Sky Survey \citep{2006AJ....131.1163S}}}
\newglossaryentry{nomad}{name=NOMAD,first={Naval Observatory Merged Astrometric Dataset \citep[NOMAD; ][]{2005yCat.1297....0Z}}, description={Naval Observatory Merged Astrometric Dataset}}
\newglossaryentry{wifes}{name=WIFES, text=\textsc{WiFeS}, first={\textsc{WiFeS} \citep{2007Ap&SS.310..255D}},  description={Wide Field Spectrograph - \gls{ifu} mounted on the 2.3\,m telescope at Siding Spring Observatory}}
\newglossaryentry{scp}{name=SCP,description={Supernova Cosmology Project, led by Saul Perlmutter}, first={Supernova Cosmology Project (SCP)}}
\newglossaryentry{hzsns}{name=HZSNS,description={High Z Supernova Search, led by Brian Schmidt}, first={High Z Supernova Search (HZSNS)}}
\newglossaryentry{vlt}{name=VLT,description={Very Large Telescope located on Cerro Paranal (Chile)}, first={Very Large Telescope (VLT)}}
\newglossaryentry{flames}{name=FLAMES,description={Multi-object, intermediate and high resolution spectrograph mounted on the  \gls{vlt}}}
\newglossaryentry{hires}{name=HIRES, description={High Resolution Echelle Spectrometer mounted on the Keck Telescope}, first={High Resolution Echelle Spectrometer \citep[HIRES;][]{1994SPIE.2198..362V}}}
\newglossaryentry{lris}{name=LRIS,description={Low Resolution Imaging Spectrometer mounted on the Keck Telescope}, first={Low-Resolution Imaging Spectrometer \citep[LRIS;][]{Oke95}}}

\newglossaryentry{essence}{name=ESSENCE,description={The `Equation of State: SupErNovae trace Cosmic Expansion' project \citep[ESSENCE;][]{2002AAS...201.7809G}}, first={`The Equation of State: SupErNovae trace Cosmic Expansion' \citep[ESSENCE;][]{2002AAS...201.7809G}}}
\newglossaryentry{ifu}{name=IFU,description={Optical instrument combining spectrographic and imaging capabilities, used to obtain spatially resolved spectra}, first={Integral Field Unit (IFU)}, firstplural={Integral Field Units (IFUs)}}

\newglossaryentry{besancon}{name=Besan\c{c}on Model, description={Model of stellar population synthesis of the Galaxy, including kinematics.}}

\newglossaryentry{int}{name=INT,description={Isaac Newton 2.5\,m Telescope}, first={Isaac Newton 2.5\,m Telescope (INT)}}
\newglossaryentry{iau}{name=IAU,description={International Astronomical Union}, first={IAU}}
\newglossaryentry{chandra}{name=Chandra,description={Chandra \xray\ Observatory (space-based)}}
\newglossaryentry{hst}{name=HST,description={Hubble Space Telescope}}
\newglossaryentry{hst.wfpc2}{name=WFPC2,description={Wide-Field Planetary Camera 2 mounted on the \gls{hst}}, first={Wide-Field Planetary Camera 2 (WFPC2)}}
\newglossaryentry{hst.acs}{name=ACS,description={Advanced Camera for Surveys mounted on the \gls{hst}}, first={Advanced Camera for Surveys (ACS)}}
\newglossaryentry{hst.wfc3}{name=WFC3,description={Wide-Field Camera 3 mounted on the \gls{hst}}, first={Wide-Field Camera 3 (WFC3)}}
\newglossaryentry{hst.cte}{name=CTE, description={charge transfer efficiency (CTE)}, first={charge transfer efficiency \citep[CTE; see ][for a description]{2009acs..rept....1C}}}

\newglossaryentry{snls}{name=SNLS,description={Supernova Legacy Survey \citep{2003AAS...203.8209P}}, first={Supernova Legacy Survey \citep[SNLS;][]{2003AAS...203.8209P}}}
\newglossaryentry{dass}{name=DASS, description={Digitized Astronomy Supernova Survey \citep{1975PASP...87..565C}}, first={Digitized Astronomy Supernova Survey \citep[DASS;][]{1975PASP...87..565C}}}
\newglossaryentry{bait}{name=BAIT, description={Berkley Automatic Imaging Telescope \citep{1993PASP..105.1164R}}, first={Berkley Automatic Imaging Telescope \citep[BAIT;][]{1993PASP..105.1164R}}}
\newglossaryentry{kait}{name=KAIT, description={Katzman Automatic Imaging Telescope \citep{2001ASPC..246..121F}}, first={Katzman Automatic Imaging Telescope \citep[KAIT;][]{2001ASPC..246..121F}}}
\newglossaryentry{loss}{name=LOSS, description={Lick Observatory Supernova Search  \citep{2000AIPC..522..103L}}, first={Lick Observatory Supernova Search \citep[LOSS;][]{2000AIPC..522..103L}}}
\newglossaryentry{ctss}{name=CTSS,description={Cal\'{a}n/Tololo Supernova Survey \citep{1993AJ....106.2392H}}, first={Cal\'{a}n/Tololo supernova survey \citep[CTSS;][]{1993AJ....106.2392H}}}
\newglossaryentry{ctio}{name= CTIO, description={Cerro Tololo Inter-American Observatory}, first={Cerro Tololo Inter-American Observatory (CTIO)}}
\newglossaryentry{ptf}{name=PTF, description={Palomar Transient Factory \citep{2009PASP..121.1334R}}, first={Palomar Transient Factory \citep[PTF;][]{2009PASP..121.1334R}}}
\newglossaryentry{batse}{name=BATSE, description={Burst and Transient Source Experiment mounted on the Compton Gamma Ray Observatory}, first={Burst and Transient Source Experiment (BATSE)}}
\newglossaryentry{bepposax}{name=BeppoSAX, description={\xray\ satellite named in honor of Giuseppe "Beppo" Occhialini}}
\newglossaryentry{rosat}{name=ROSAT, description={short for R\"{o}ntgensatellit}, first={ROSAT}}
\newglossaryentry{hete2}{name=HETE2, description={High Energy Transient Explorer}, first={High Energy Transient Explorer (HETE)}}
\newglossaryentry{ska}{name=SKA, description={Square Kilometre Array}, first={Square Kilometre Array (SKA)}}

\newglossaryentry{gnirs}{name=GNIRS, description={Gemini Near InfraRed Spectrograph mounted on the Gemini North Telescope}}
\newglossaryentry{gmosn}{name=GMOS, description={Gemini Multi Object Spectrograph mounted on the
 Gemini North Telescope}, first={GMOS \citep[Gemini Multi Object Spectrograph;][]{2004PASP..116..425H}}}
\newglossaryentry{swift}{name=Swift, description={Swift Gamma-Ray Burst Mission}}
\newglossaryentry{vla}{name=VLA, description={Very Large Array radio telescope located in North America}, first={Very Large Array (VLA)}}
\newglossaryentry{evla}{name=EVLA, description={Extended Very Large Array radio telescope located in North America}, first={Extended Very Large Array (EVLA)}}
\newglossaryentry{sdss}{name=SDSS, description={Sloan Digital Sky Survey}}
\newglossaryentry{dss}{name=DSS, description={Digitized Sky Survey}}
\newglossaryentry{skymapper}{name=SkyMapper, description={SkyMapper telescope \citep{2007PASA...24....1K}}, first={SkyMapper \citep{2007PASA...24....1K}}}
\newglossaryentry{panstarrs}{name=PanSTARRS, description={Panoramic Survey Telescope \& Rapid Response System \citep{2004SPIE.5489...11K}}, first={Panoramic Survey Telescope \& Rapid Response System \citep[PanSTARRS;][]{2004SPIE.5489...11K}}}
\newglossaryentry{lsst}{name=LSST, description={Large Synoptic Survey Telescope}, first={Large Synoptic Survey Telescope \citep[LSST;][]{2006AAS...209.8604P}}}
\newglossaryentry{ppmxl}{name=PPMXL, description={PPMXL Catalog of Positions and Proper Motions on the ICRS \citep{2010AJ....139.2440R}}}
\newglossaryentry{gaia}{name=GAIA, description={Global Astrometric Interferometer for Astrophysics \citep{2001A&A...369..339P}}, first={Global Astrometric Interferometer for Astrophysics \citep[GAIA;][]{2001A&A...369..339P}}}
\newglossaryentry{ligo}{name=LIGO, description={Laser Interferometer Gravitational Wave Observatory}, first={Laser Interferometer Gravitational Wave Observatory \citep[LIGO;][]{1992Sci...256..325A}}}
\newglossaryentry{aligo}{name=Advanced LIGO, description={Advanced LIGO}, sort=ligo2}
\newglossaryentry{lisa}{name=LISA, description={Laser Interferometer Space Antenna \citep{1994ESAJ...18..219J}}, first={Laser Interferometer Space Antenna \citep[LISA;][]{1994ESAJ...18..219J}}}
\newglossaryentry{ccd}{name=CCD,description={Charged Coupled Device}, first={charged coupled device (CCD)}, firstplural={charged coupled devices (CCDs)}}

\newcommand{\sn}[2]{SN~#1#2\xspace}

\newglossaryentry{irc}{name=IRC, text={IRC}, description={infrared catastrophe}, first={infrared catastrophe \citep[IRC;][]{1980PhDT.........1A}}}

\newglossaryentry{sn}{name=Supernova, text={SN}, plural={SNe}, description={exploding star}, nonumberlist=true, first={supernova (SN)}, firstplural={supernovae (SNe)}}
\newglossaryentry{snia}{name=Type~Ia (SN~Ia), text={SN~Ia}, description={Thermonuclear explosion of a white dwarf - spectra show no hydrogen but a strong silicon line},first={Type~Ia supernova (SN~Ia)}, firstplural={Type Ia supernovae (SNe~Ia)}, plural={SNe~Ia}, parent=sn, nonumberlist=true}
\newcommand{\sneia}{\glspl*{snia}\xspace}

\newglossaryentry{branchnormal}{name={branch-normal}, text=\textit{Branch-normal}, description={Large homogeneous class of Type Ia Supernovae, defined in \citet{1993AJ....106.2383B}}, first={\textit{Branch-normal} SNe Ia \citep{1993AJ....106.2383B}}, parent=snia} 
\newglossaryentry{91t}{name={91T-like}, description={Luminous class of Type Ia supernovae similar to \sn{1991}{T} \citep{1992AJ....103.1632P}} , first={91T-like}, parent=snia} 
\newglossaryentry{91bg}{name={91bg-like}, description={Faint class of Type Ia supernovae similar to \sn{1991}{bg} \citep{1992AJ....104.1543F}}, first={91bg-like}, parent=snia} 
\newglossaryentry{02cx}{name={02cx-like}, description={Peculiar class of Type Ia supernovae similar to \sn{2002}{cx} \citep{2003PASP..115..453L}}, first={02cx-like \sneia\ \citep{2003PASP..115..453L}}, parent=snia} 

\newglossaryentry{snibc}{name=Type~Ib/c, text={SN~Ib/c}, description={Collapse of the core of a massive star -  spectrum shows no hydrogen and no silicon line},first={Type~Ib/c supernova (SN~Ib/c)}, firstplural={Type~Ib/c supernovae (SNe~Ib/c)}, plural={SNe~Ib/c}, parent=sn}

\newglossaryentry{snib}{name=Type~Ib, text={SN~Ib}, description={Spectrum shows no hydrogen and no silicon, but helium line},first={Type Ib supernova (SN~Ib)}, firstplural={Type~Ib supernovae (SNe~Ib)}, plural={SNe~Ib}, parent=snibc}

\newglossaryentry{snic}{name=Type~Ic, text={SN~Ic}, description={Spectrum shows no hydrogen, no silicon and no helium line},first={Type~Ic supernova (SN~Ic)}, firstplural={Type~Ic supernovae (SNe~Ic)}, plural={SNe~Ic}, parent=snibc}

\newglossaryentry{snii}{name=Type~II, text={SN~II}, description={Collapse of the core of a massive star - spectrum shows strong hydrogen line},first={Type~II supernova (SN~II)}, firstplural={Type~II supernovae (SNe~II)}, plural={SNe~II}, parent=sn}

\newglossaryentry{sniib}{name=Type~IIb, text={SN~IIb}, description={Spectrum shows hydrogen and helium lines},first={Type~IIb supernova (SN~IIb)}, firstplural={Type~IIb supernovae (SNe~IIb)}, plural={SNe~IIb}, parent=snii}

\newglossaryentry{sniip}{name=Type~II~Plateau (Type IIP), text={SN~IIP}, description={Lightcurve shows plateau},first={Type~IIP supernova (SN~IIP)}, firstplural={Type~II Plateau supernovae \citep[SNe~IIP;][]{1979A&A....72..287B}}, plural={SNe~IIP}, parent=snii}

\newglossaryentry{sniil}{name=SN~II~Linear, text={SN~IIL}, description={Lightcurve shows no plateau, but linear decline},first={Type~IIL supernova (SN~IIL)}, firstplural={Type~II~Linear supernovae \citep[SNe~IIL;][]{1990MNRAS.244..269S}}, plural={SNe~IIL}, parent=snii}

\newglossaryentry{sniin}{name=Type II narrow-lined (Type IIn), description={Spectrum shows narrow lines},first={Type~II~narrow-lined supernova (SN IIn)}, firstplural={Type~IIn supernovae (SNe~IIn)}, plural={SNe~IIn}, parent=snii}

\newglossaryentry{snr}{name=Remnant (SNR), text=SNR, description={Remnant left visible post-explosion}, first={supernova remnant (SNR)}, firstplural={supernova remnants (SNRs)}, parent=sn}

\newglossaryentry{dtd}{name=DTD,description={delay time distribution - expected supernova rate over time after a brief outburst of starformation},first={delay time distribution (DTD)}, firstplural={delay time distributions (DTDs)}, plural=DTDs}

\newglossaryentry{hvg}{name=HVG,description={high velocity gradient - Type Ia supernovae with a fast evolution of photospheric velocity},first={high velocity group (HVG)}, firstplural={high velocity groups (HVGs)}, plural=HVGs, parent=snia}

\newglossaryentry{lvg}{name=LVG,description={low velocity gradient - Type Ia supernovae with a slow evolution of photospheric velocity},first={low velocity group (LVG)}, firstplural={low velocity groups (LVGs)}, plural=LVGs, parent=snia}

\newglossaryentry{wd}{name=white dwarf (WD), text=WD, description={White Dwarf - extremely dense stellar remnant}, first={white dwarf (WD)}}
\newglossaryentry{onemgwd}{name= Oxygen/Neon (ONe), text={ONe-WD},description={Oxygen/Neon White Dwarf}, first={oxygen/neon White Dwarf (ONe-WD)}, parent=wd}
\newglossaryentry{cowd}{name=carbon/oxygen (CO), text={CO-WD}, description={carbon/oxygen white dwarf}, first={carbon/oxygen white dwarf (CO-WD)}, firstplural = {carbon/oxygen white dwarfs (CO-WDs)}, parent=wd}

\newglossaryentry{sds}{name=SD-Scenario,description={single-degenerate scenario (single white dwarf accreting from non-degenerate companion)}, first={single-degenerate scenario (SD-scenario)}}

\newglossaryentry{dds}{name=DD-Scenario, description={double degenerate scenario (merging of two white dwarfs)}, first={double-degenerate scenario (DD-scenario)}}

\newglossaryentry{sss}{name=SSS, text={supersoft \xray\ source}, description={supersoft \xray\ source - believed to be emitted by nuclear fusion on a white dwarf's surface}}

\newglossaryentry{amcvn}{name=AM CVn, description={AM Canum Venaticorum star (white dwarf accreting hydrogen poor matter from a companion star; see \cite{2005ASPC..330...27N})}}

\newglossaryentry{rlof}{name=RLOF, description={Roche Lobe Overflow (see \citet{1971ARA&A...9..183P} for a more detailed description)}, first={Roche-lobe overflow (RLOF)}}

\newglossaryentry{mchan}{name={Chandrasekhar mass~}, text={Chandrasekhar~mass}, symbol={\ensuremath{M_\textrm{Chan}}}, plural={Chandrasekhar~masses}, description={Mass when the core of a star collapses due to insufficient degeneracy pressure - for a white dwarf $\approx1.38\,M_\odot$ see \citet{1931ApJ....74...81C}}, first={Chandrasekhar~mass \citep[$M_\textrm{Chan}=1.38\,M_\odot$;][]{1931ApJ....74...81C}}, sort=mchan}

\newglossaryentry{w7}{name={W7 model},description={W7 model \citep{1984ApJ...286..644N}},first = {W7 model \citep{1984ApJ...286..644N}}}

\newglossaryentry{ew}{name=Equivalent Width, text={EW}, description={width of a rectangle that has the same area as a spectral line when taken to zero flux}, first={equivalent width (EW)}, firstplural={equivalent widths (EWs)}}
\newglossaryentry{agb}{name=AGB,description={Asymptotic Giant Branch}, first={Asymptotic Giant Branch (AGB)}}
\newglossaryentry{cmb}{name=CMB,description={Cosmic Microwave Background}}
\newglossaryentry{csm}{name=CSM,description={Circumstellar Medium}, first={circumstellar medium (CSM)}}
\newglossaryentry{csi}{name=CSI,description={Circumstellar Interaction}, first={circumstellar interaction (CSI)}}
\newglossaryentry{ism}{name=ISM,description={Interstellar Medium}, first={interstellar medium (ISM)}}
\newglossaryentry{ige}{name=IGE,description={Iron Group Element}, first={iron group element (IGE)}, firstplural={iron group elements (IGEs)}}
\newglossaryentry{epm}{name=EPM,description={Expanding Photosphere Method \citep{1974ApJ...193...27K}}, first={Expanding Photosphere Method (EPM)}}
\newglossaryentry{aic}{name=AIC,description={Accretion Induced Collapse}, first={accretion induced collapse (AIC)}}
\newglossaryentry{ime}{name=IME,description={Intermediate Mass Element}, first={intermediate mass element (IME)}, firstplural={intermediate mass elements (IMEs)}}
\newglossaryentry{h0}{name=\ensuremath{H_0},description={Hubbles constant}}
\newglossaryentry{nse}{name=NSE,description={Nuclear Statistical Equilibrium}, first={nuclear statistical equilibrium (NSE)}}
\newglossaryentry{cdm}{name=CDM,description={Cold Dark Matter}}
\newglossaryentry{grb}{name=GRB,description={Gamma Ray Burst}, first={Gamma Ray Burst (GRB)}, firstplural={Gamma Ray Bursts (GRBs)}}
\newglossaryentry{donor}{name=donor,description={non-degenerate companion in the \gls{sds}}}
\newglossaryentry{mainsequence}{name=main sequence,description={main sequence star}}
\newglossaryentry{redgiant}{name=red giant,description={red giant star}}
\newglossaryentry{mlcs}{name=MLCS,description={Multicolor Light Curve Shape method \citep[MLCS;][]{1996ApJ...473...88R}}, first={Multicolor Light-Curve Shape method \citep[MLCS;][]{1996ApJ...473...88R}}}
\newglossaryentry{rsoph}{name=RS~Ophiuci ,description={white dwarf accreting from a red giant - assumed progenitor of the \gls{sds}}, sort=rsoph}
\newglossaryentry{usco}{name=U~Scorpii,description={white dwarf accreting from a main sequence star - assumed progenitor of the \gls{sds}}, sort=usco}
\newglossaryentry{rcw86}{name=RCW~86,description={supernova remnant sometimes associated with \sn{185}{}}, sort=rcw86}
\newglossaryentry{casa}{name=Cas~A,description={Cassiopeia A supernova remnant - probably a \gls{snib} event}}
\newglossaryentry{cepheid}{name=Cepheid,description={very luminous variable star with a strong luminosity period relationship}}
\newglossaryentry{urca}{name=Urca, text=\textit{Urca}, description={process predominatly contributing to cooling in stars. The \textit{Urca} process consists of alternating electron-capture and $\beta^{-}$ decay of two nuclei pairs.},sort=urca} 
\newglossaryentry{alphacen}{name=Alpha Centauri,description={one of the brightest stars in the night sky and a close binary}}
\newglossaryentry{pcygni}{name={P Cygni}, text={P Cygni},description={a hypergiant luminous blue variable with strong winds. Often referred to as a description for their line profiles showing a emission peak at the rest wavelength of the line and a blue-shifted absorption trough.}}

\newglossaryentry{teff}{name={effective temperature~}, text={effective temperature}, symbol={\ensuremath{T_\textrm{eff}}}, description={Temperature of a blackbody emitting the same total energy}, sort=teff}

\newglossaryentry{logg}{name={surface gravity~}, text={surface gravity}, symbol={\ensuremath{\textrm{log}\,g}}, description={gravity at the surface of a star}, sort=logg}
\newglossaryentry{feh}{name={metallicity~}, text={metallicity}, symbol=\textrm{[Fe/H]},description={iron abundance relative to the sun}, sort=feh}

\newglossaryentry{texp}{name={time since explosion~}, text={time since explosion}, text={time since explosion}, symbol={\ensuremath{t_{\rm exp}}},description={time since explosion (measured in days)}, sort=texp, first={time since explosion (\ensuremath{t_{\rm exp}})}}

\newglossaryentry{lmc}{name=LMC,description={Large Magellanic Cloud}, first={Large Magellanic Cloud (LMC)}, sort=lmc}
\newglossaryentry{smc}{name=SMC,description={Small Magellanic Cloud}, sort=smc}
\newglossaryentry{z}{name=\ensuremath{z},description={redshift}, sort=z}

\submitjournal{PASP}
\usepackage{graphicx}
\usepackage{txfonts}
\newcommand{\arxiv}{\textit{arXiv}\xspace}

\begin{document} 

\title{Knowledge discovery through text-based similarity searches for astronomy literature}

\author{Wolfgang~E.~Kerzendorf}

\address{European Southern Observatory, Karl-Schwarzschild-Strasse 2, 85748 Garching, Germany}
  \begin{abstract}The increase in the number of researchers coupled with the ease of publishing and distribution of scientific papers (due to technological advancements) has resulted in a dramatic increase in astronomy literature. This has likely led to the predicament that the body of the literature is too large for traditional human consumption and that related and crucial knowledge is not discovered by researchers. In addition to the increased production of astronomical literature, recent decades have also brought several advancements in computational linguistics. Especially, the machine-aided processing of literature dissemination might make it possible to convert this stream of papers into a coherent knowledge set. In this paper, we present the application of computational linguistics techniques to astronomy literature. In particular, we developed a tool that will find similar articles purely based on text content from an input paper. We find that our technique performs robustly in comparison with other tools recommending articles given a reference paper (known as recommender system). Our novel tool shows the great power in combining computational linguistics with astronomy literature and suggests that additional research in this endeavor will likely produce even better tools that will help researchers cope with the vast amounts of knowledge being produced.
 \end{abstract}
  

%
\section{Introduction}

Since the inception of writing, human knowledge has steadily increased
as has the number, and size, of published works.
The output of the scientific community has doubled every nine years
over the past decades \citep{bornmann2015growth}. 

The computing and internet revolution has made the publication and
dissemination of these works easy and with the advent of open access
channels pluralistic. The public repository \arxiv
has provided open access to almost the entire corpus of publications since 
1992 in the physical sciences . 

Given the rise of publications each year and the fixed capacity of a human
to process information either we shall narrow the specialization range in
each field to limit the breadth of the necessary knowledge base or have
new tools that filter the available publications. 

In astronomy, the \gls{ads} has provided access (in addition to many other accomplishments such as digitizing old articles) to this large amount of literature with a search interface that captures the traditional way of accessing information (name of first author and year) extremely well. Newer iterations of this system \citep{2015ASPC..495..401C} have started to branch out and allow not only search algorithms but provide certain bibliometric statistics as well as a recommender system (named ``Suggested Articles''). This recommender system is based on citations, text similarity, and co-readership \citep[as described on the ADS 2.0 website and suggested in][]{2010APS..MAR.S1244H, 2011ASSP...24...23K}. Such recommender systems will be a first step to tackle a world in which the scientific literature has massively outgrown the memory capacity of human brains. 

In this paper, we present a new method for article recommendations starting from
a reference article or text. We employ the techniques of text similarity and
specifically avoid citations. This strict abstention from citation was chosen due to the fact that citations are influenced by many factors and may not provide an unbiased
link between publications \citep[several examples in][]{vanWesel2014}. A web service, based on the presented tools and techniques, can be found at \url{http://opensupernova.org/deepthought}.

In Section~\ref{sec:data_proc}, we describe the data acquisition, initial vetting, and processing. Section \ref{sec:method},
describes the method used and some statistics arising. An overview of the
framework used in this work and its application to several example papers in Section 4. Caveats and possible improvements are discussed in Section 5 and
we conclude with an outlook to the future in Section 6. 

\section{Data Processing}
\label{sec:data_proc}
For our initial raw corpus, we considered all papers submitted to \arxiv. Using the bulk data access\footnote{e.g. \url{s3://arxiv/src/arXiv_src_1001_001.tar}}, we downloaded the entire corpus. After a series of operations (discarding any non-latex submissions), we arrived at individual source directories (a total of \num{1301668}).
This work focuses currently on the field of astronomy. For all entries, we harvested the metadata through the OAI protocol for metadata harvesting (OAI-PMH) and then selected all \arxiv entries that had \textit{astro} in the \textit{subject} node of the metadata. This amounts to a total corpus size of \num{232680} papers. 

In each source compilations, we tried to identify the main tex file by requiring a single valid \textit{\textbackslash begin\{document\}} clause and processed this further with \textsc{latexpand}\footnote{\url{https://www.ctan.org/pkg/latexpand}} to a single document that would contain all relevant text content. Not all entries had a uniquely identifiable main \textit{tex-file}. 

The resulting \textit{tex-files} were further processed removing the most common environments:

\begin{itemize}
\item \textrm{Figures}\\
\textit{figure, picture}
\item \textrm{Tables}\\
\textit{table, deluxetable}
\item Equations\\
\textit{equation, align, subequations, eqnarray, array, matrix}
\end{itemize}

Then we removed any text before the first section command or if this was not present any text before end abstract. 

The final step of the raw reduction process was the removal of latex commands using the \textsc{opendetex} software \footnote{\url{https://github.com/pkubowicz/opendetex}}. 

\begin{figure*}
  \centering
  \includegraphics[width=\textwidth]{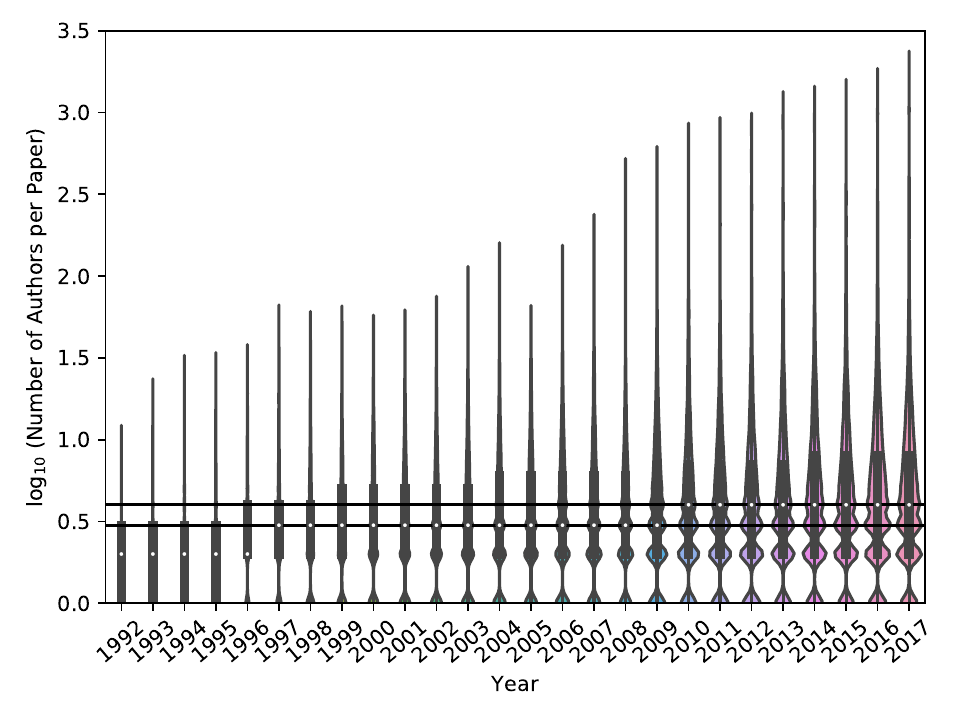}
\caption{Kernel density estimate of the distribution of the number of authors in each year (using a kernel bandwidth of 0.1). The width of the distribution scales with total number of papers for each year. The black lines indicate the median of 3 and 4 authors.}
  \label{fig:no_authors}
\end{figure*}

These unprocessed data and metadata already allow for some interesting statistics to study publication behavior. As \arxiv has already shown\footnote{see \url{https://arxiv.org/help/stats/2016_by_area/index}} there is a roughly linear increase in astronomy papers. The number of authors also is an important metric to understand the culture of a field. Here often people turn to the average number of authors as a measure \citep[e.g.][]{10.15200/winn.141832.26907}. While this is a useful statistics, it is heavily biased towards the few papers with an extreme number of authors. 
Figure~\ref{fig:no_authors} shows that the general number of authors remains relatively low but there is a very small number of papers with an increasingly extreme number of authors (only 0.5\% of papers have more than one hundred authors).

\subsection{Natural Language Processing}
These raw texts are ready for natural language processing and the following steps use the \gls{nltk} tools extensively. 

The first step in this process is to break the text into individual words using \textsc{nltk.tokenize\_words}, which splits into individual words and removes all punctuation - except the period (which is treated as a single word at this step). 

The next process is to remove stop words such as \textit{these, those, am, is, are} (using the English stop words defined in \textsc{nltk.corpus.stopwords}). 

The final step of processing is to lemmatize the words, which is the process of grouping together the different inflected forms of a word so it can be analyzed as a single item. 
For this process we use the tool \textsc{nltk.wordnet.morphy} to bring the words back to their original forms (galaxies maps to galaxy, expanding maps to expand, etc.). We discard words that do not have a corresponding entry in the dictionary provided \citep[\textsc{WordNet;}][]{fellbaum1998wordnet}. 

After this final step, we are left with a corpus consisting of \num{201997} documents. 

\section{Method}
\label{sec:method}
In this work, we will entirely rely on the bag-of-words technique \citep[][]{harris1954distributional}, that disregards grammar and word order, treating the document just as a collection of words. The features that we use for our analysis are several statistics based on word frequency. For all feature extraction tasks, we relied on \gls{scikit-learn}. 

The first step for any of these methods is building a vocabulary of unique words. This is helped by the fact that we transformed our document removing any stop words and transforming the words to their simplest form. 

This vocabulary consists of $\approx ~\num{32000}$ words. This is only slightly larger than the $\approx\num{20000}$ \citep{goulden1990large} words a well educated native speaker knows and much lower than the $\approx\num{170000}$ words in the Oxford English dictionary \citep{simpson1989oxford}.  
Our vocabulary can then be used to vectorize (feature extract) the documents to vectors the size of the dictionary (in our case $\approx\num{32000}$).

The simplest case is the use of a binary statistics for feature extraction which will only encode if a word is present or not present. This statistics can give a rough overview of the content but will de-weight more frequently used words and thus possible shift the inferred topic of the document in statistical analysis. 

\begin{figure}
  \centering
  \includegraphics[width=\columnwidth]{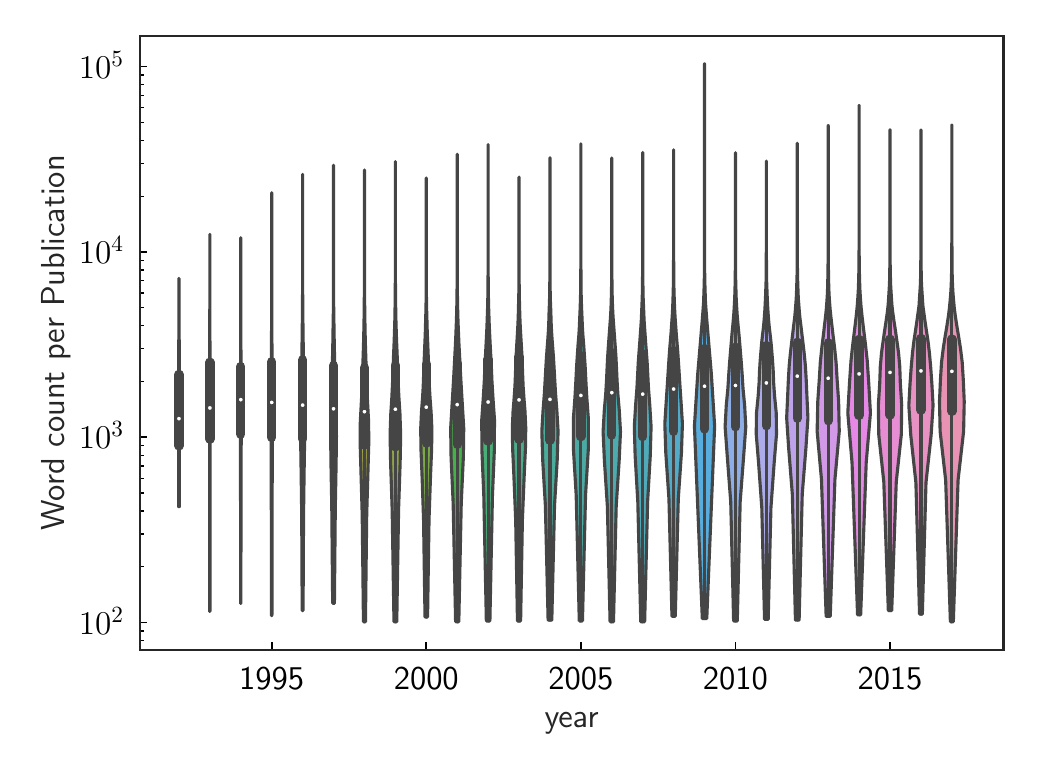}

\caption{Distribution of word count without stop words for all papers published (to astro-ph) in each year}
  \label{fig:median_word_per_paper}
\end{figure}

\begin{figure}
  \centering
  \includegraphics[width=\columnwidth]{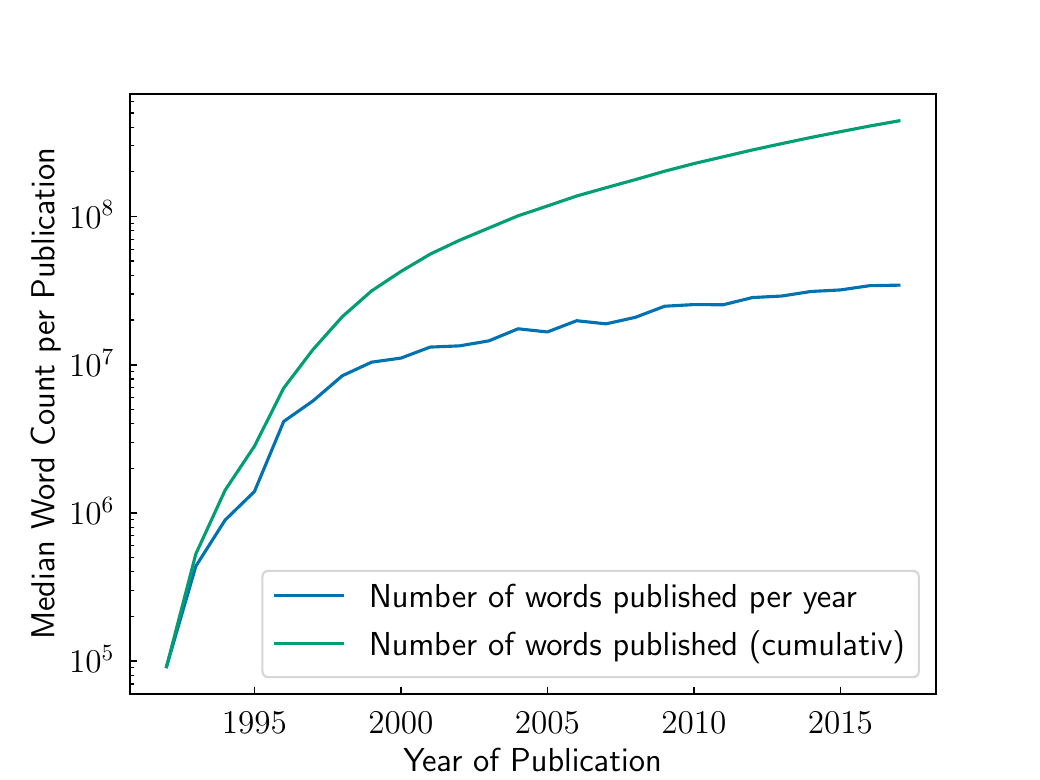}

\caption{Total Number of words without stop words for all papers published (to astro-ph) in each year.}
  \label{fig:words_per_year}
\end{figure}

The first statistics we have performed on the corpus of documents is a simple count vectorization (using \textsc{scikit-learn.feature\_extraction.CountVectorizer}). 
This count measure allows us to quantify the growth of literature since the conception of \arxiv. Figure~\ref{fig:median_word_per_paper} shows, that while the growth in document size (number of words using the processed document word counts as a proxy) has been maybe a factor of 1.5 over the last decade, astronomical literature in total has grown exponentially  (see Figure~\ref{fig:words_per_year}) over the same period. This given measure of ``word count'', however, has several drawbacks, the most important one being that it increases with document length and thus does not give a useful measure for word importance. 

The last vectorization technique is a natural extension of the word counts that aims at emphasizing a word's importance in a text - thus ideal for assessing the content of a paper. In the following, we will use the moniker ``term'' and ``word'' interchangeably as our analysis uses only one-word terms (unigrams). The term frequency $tf(t,d)$ method normalizes the simple word count by the number of words in the document. This relies on the assumption that the importance (or weight) of a term in documents is proportional to this term frequency \citep{5392697}. In addition to term frequency, we want to quantify the information content a specific term carries. \citet{sparck1972statistical} have introduced the concept of inverse document frequency $idf(t, d)$. We use the inverse document frequency given in \textsc{scikit-learn} as $\log{\frac{1 + n_d}{1+df(d, t)}}$, where $n_d$ is the total number of documents and $df(d, t)$ is the number of documents containing the given term. The combination of both measures gives the well established $tf(t, d) \times idf(t,d)$ (henceforth TFiDF) measure which weights terms highly that have a high information content due to their rarity. This measure is used in several machine learning tasks including finding similar texts.

\section{Similarity in papers}
\label{sec:similarity}
We explore how the TFiDF statistics can lead to knowledge discovery. 

For this purpose, we first normalize our document vectors using the euclidean norm $\vec{d}_\textrm{norm} = \frac{\vec{d}}{||\vec{d}||_2}$ before proceeding further. Leaving us with the entire sparse matrix (which is available upon request). Here, we present an example of such a matrix (with the different document vectors as rows and the columns representing different words/terms):
\begin{equation}
A_\textrm{TFiDF} = 
\bordermatrix{~ & star & model & \cdots & galaxy\cr
              \textrm{arXiv}-1 & 0.021 & \cdots & 0 \cr
              \textrm{arXiv}-2 & 0 & 0.03 & \cdots & 0\cr
              \vdots           & 0.019 & 0.016 & \cdots & 0 \cr
              \textrm{arXiv}-n & 0 & 0 & \cdots & 0.023 \cr}
\end{equation}
We simply use the cosine distance by choosing a document that we want to compare and multiplying this with the TFiDF matrix $\vec{v}_\textrm{similarity} = A_\textrm{TFiDF} \times \vec{d}_\textrm{norm}$ to measure text similarity. An example service that showcases this technique is available at 
\url{http://opensupernova.org/deepthought}.

\subsection{Example: SN 1006 companion search}
In the following, we will use this approach on a paper that is well known to the author: ``Hunting for the Progenitor of SN 1006: High-resolution Spectroscopic Search with the FLAMES Instrument'' \citep{2012ApJ...759....7K}. This paper describes the failed attempt to find a surviving companion star (often called donor star) to a supernova (likely caused by a white dwarf) searching in one of the supernova remnants in our Galaxy. 

We first measure the most important words with the presented algorithm. For this purpose, we inverse sort our $\vec{v}_\textrm{similarity}$ and use the first best 100 matches. We multiply the $\vec{d}_\textrm{norm}$ and look for the highest entries in the resulting vector which should give us the most important words that the algorithm matches. Figure~\ref{fig:example_papers} clearly shows that very relevant words that one would expect when writing a paper about searching for a donor star in a supernova remnant likely caused by a white dwarf.

This might also be obtained by doing the typical manual search technique using the citations to the current paper as well as the references to the current paper. However, we aim to find papers that would not be found using this common technique. 

\begin{figure}
  \centering
  \includegraphics[width=\columnwidth]{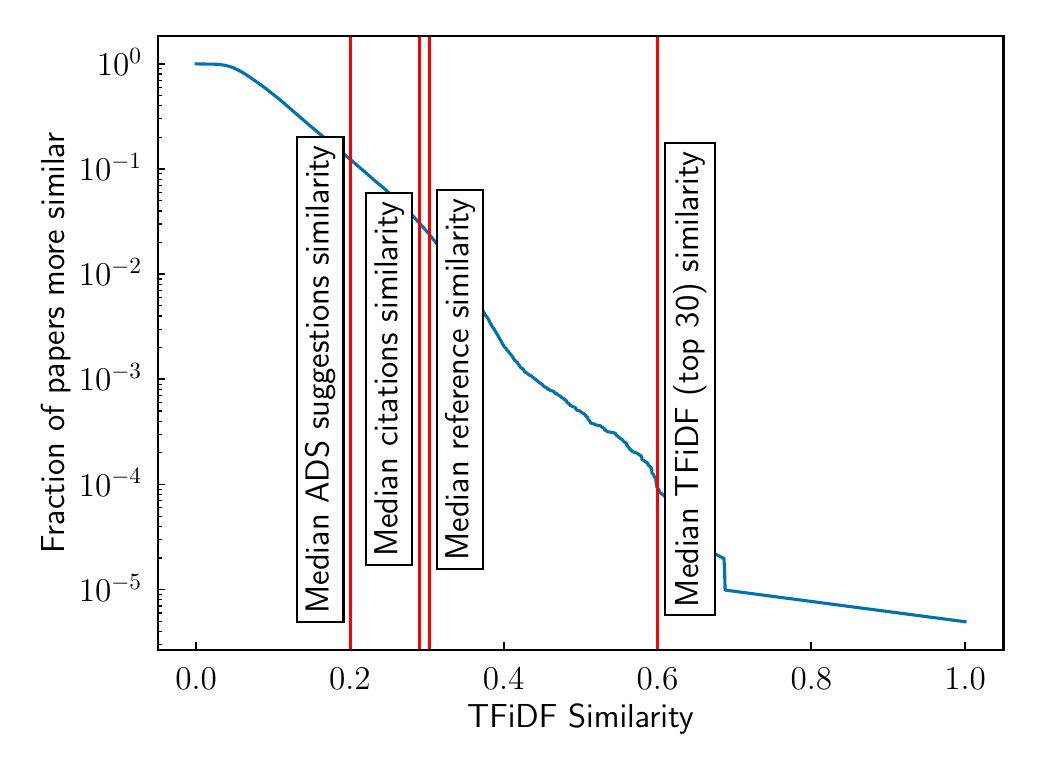}
 
\caption{Comparing cumulative similarity to the given TFiDF similarity number. For example, papers that have a higher similarity than the median similarity of the references only make up 5\% of the entire corpus of papers.}
  \label{fig:group_simil_compare}
\end{figure}

All references to our test paper (12 in total found in our \arxiv corpus) have a median similarity of 0.38. Only 3\% of papers (see Figure~\ref{fig:group_simil_compare}) in the astronomy corpus are more similar than this median which suggests that as expected the citations are highly relevant. 

All citations to our test paper (30 in total were found in our \arxiv corpus) have a median similarity of 0.26 and 5\% of papers are more similar than this suggesting that the citations to this article come from a more varied field than the original references (or that references had been forgotten).

Next, we test if the algorithm's top 30 papers  (similar to the citations) are in neither citations nor references and compare this to the relevance of the other papers. Figure~\ref{fig:example_papers} shows the comparison between the median cosine distance of the cited papers, the median cosine distance between the references and the median top 30 results excluding papers from both of these groups. This demonstrates that such a system can find relevant papers that could easily be missed otherwise. 

ADS has recently implemented a similar system \citep{2015ASPC..495..401C}, however we find that it does not give as relevant matches as the presented algorithm. There are 30\% of papers that are more similar to the document in question compared to the median similarity of the suggested papers. Manual inspection also shows that some of the suggested papers (e.g. ``Maps of Dust Infrared Emission for Use in Estimation of Reddening and Cosmic Microwave Background Radiation Foregrounds'' by 
\citealt{1998ApJ...500..525S}) might be very broadly related but not very relevant. The precise algorithm to determine the suggested papers for ADS is not accessible thus prohibiting a quantitative comparison.

\subsection{Example: Globular cluster} 
\begin{figure*}
  \centering
  \includegraphics[width=\columnwidth]{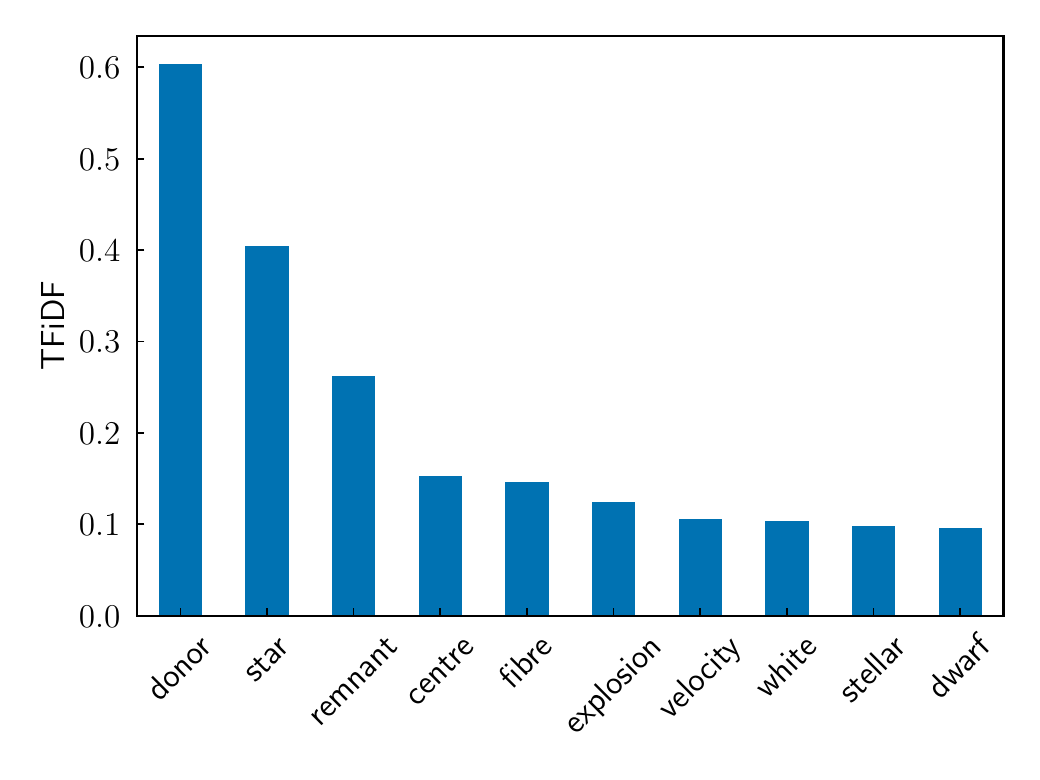}
  \includegraphics[width=\columnwidth]{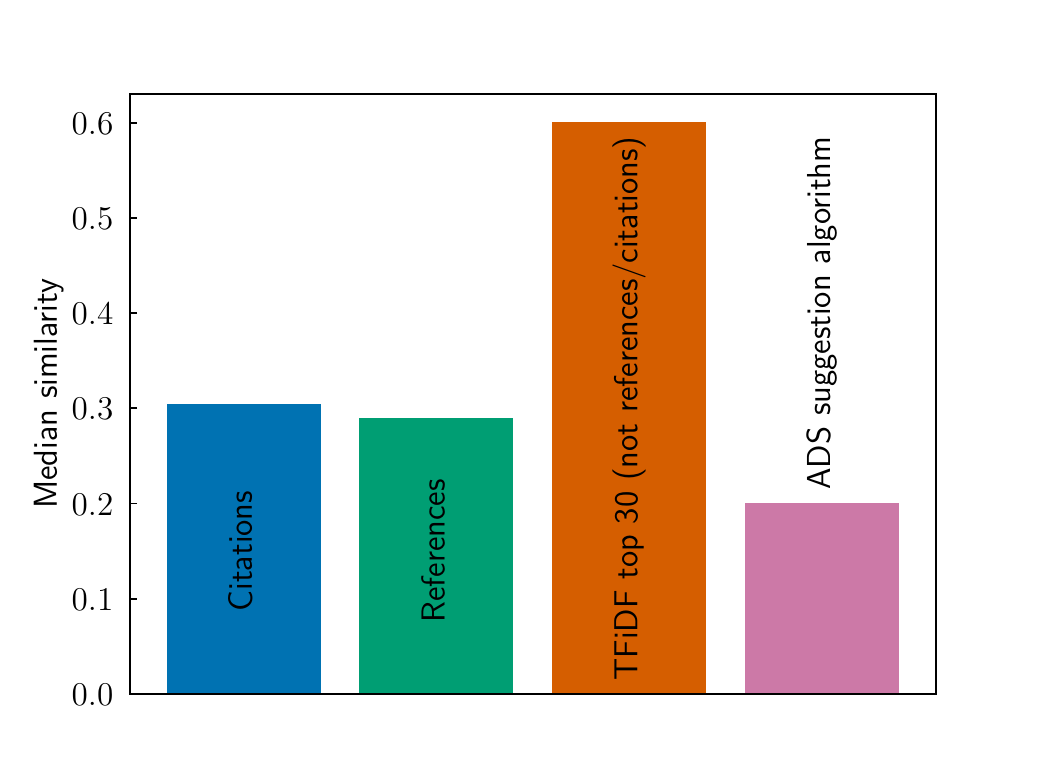}
  
  \includegraphics[width=\columnwidth]{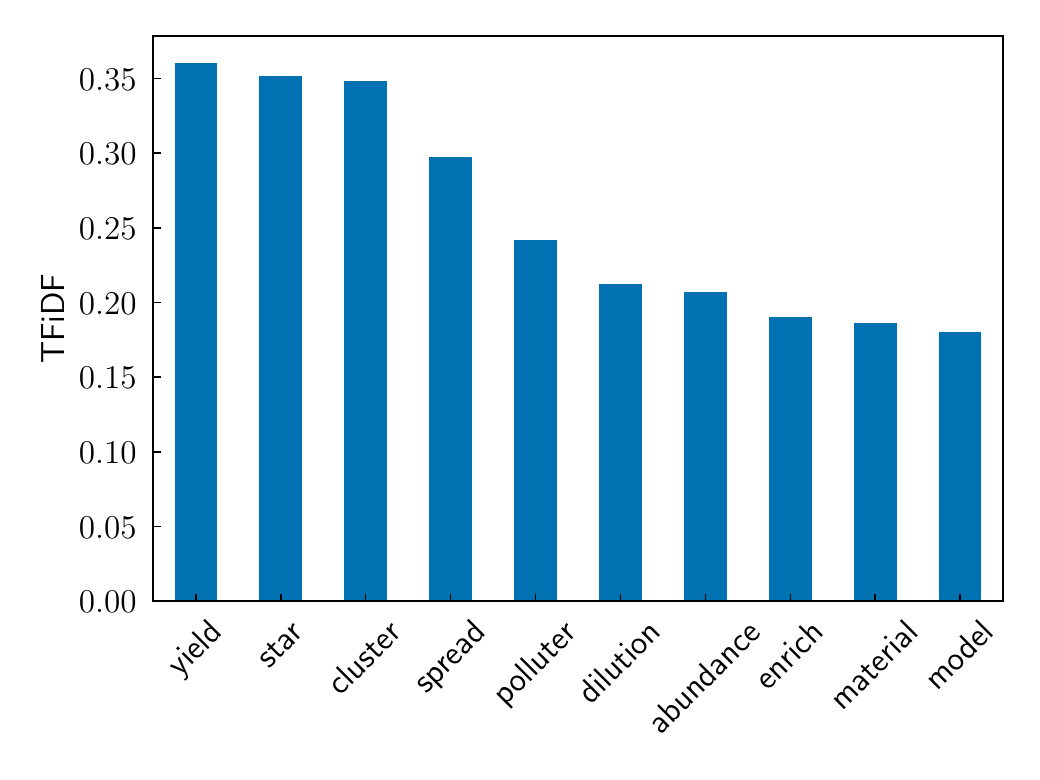}
\includegraphics[width=\columnwidth]{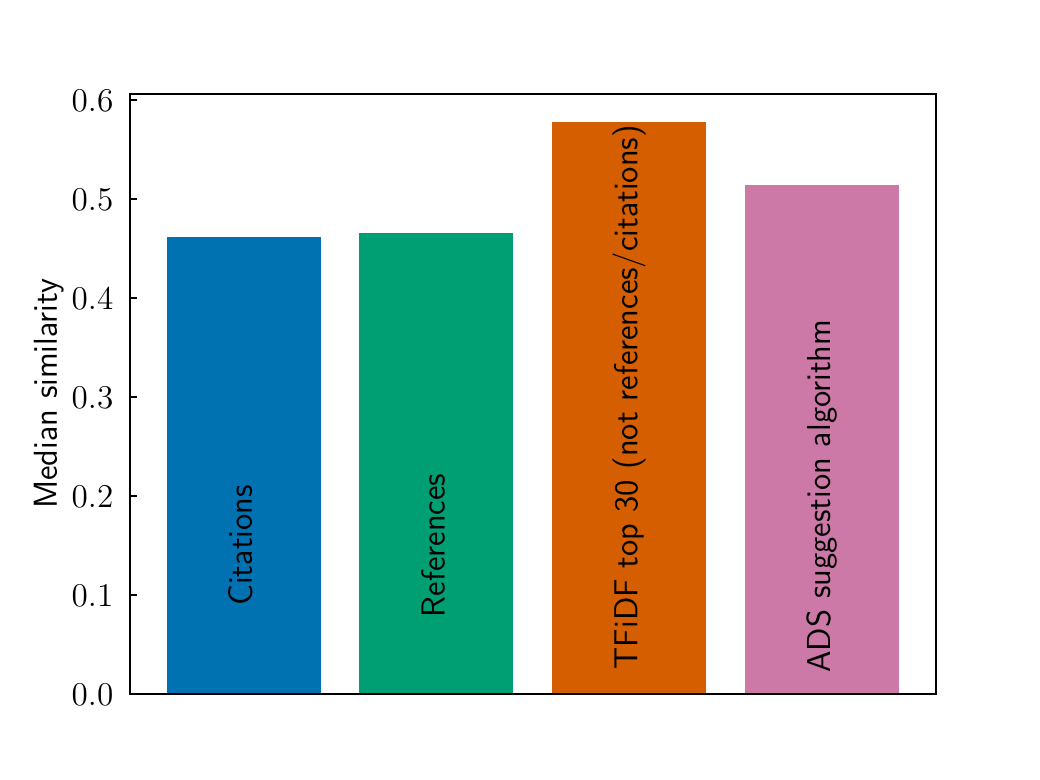}
  \includegraphics[width=\columnwidth]{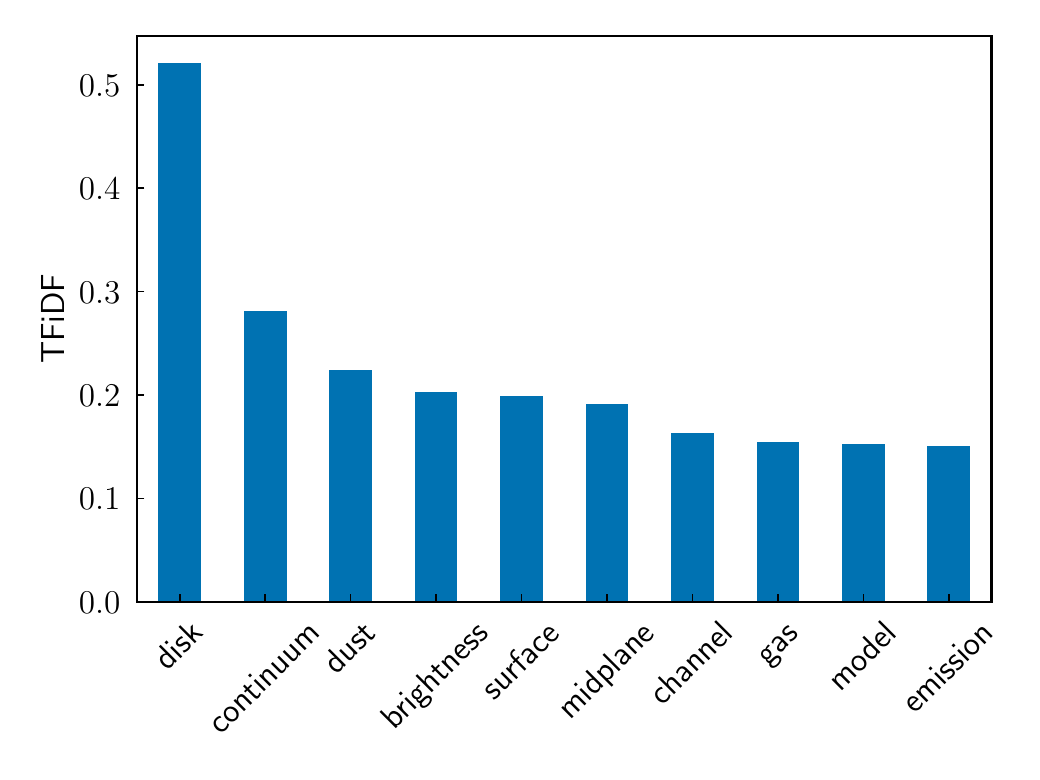}
\includegraphics[width=\columnwidth]{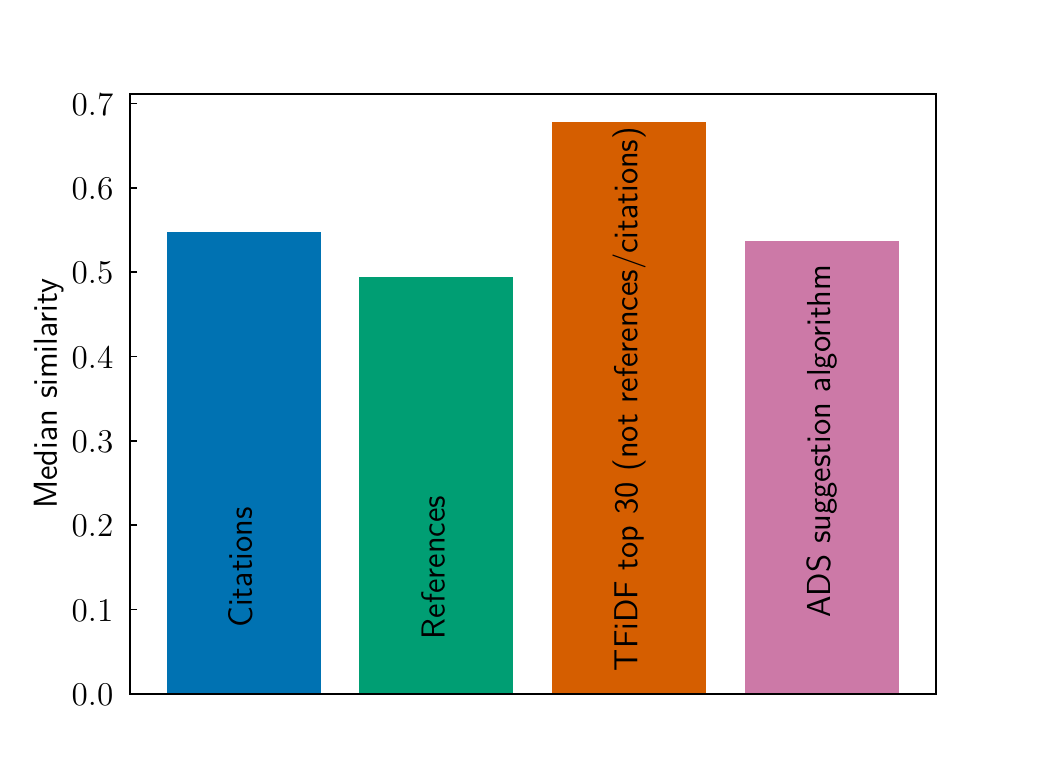}
\caption{The TFiDF method applied to three distinct papers (top) ``Hunting for the Progenitor of SN 1006: High-resolution Spectroscopic Search with the FLAMES Instrument '' \citep{2012ApJ...759....7K} (middle) ``A general abundance problem for all self-enrichment scenarios for the origin of multiple populations in globular clusters'' \citep{2015MNRAS.449.3333B} (bottom) ``Unveiling the gas-and-dust disk structure in HD 163296 using ALMA observations'' \citep{2013A&A...557A.133D}.
The left plot shows the TFiDF weight for the ten most words with the highest weights while the right plot in each shows the similarity metric applied to several collections associated with these papers.}
  \label{fig:example_papers}
\end{figure*}

The second test paper we use is ``A general abundance problem for all self-enrichment scenarios for the origin of multiple populations in globular clusters'' by \citet{2015MNRAS.449.3333B}. This paper points out a possible flawed explanation for abundance anomalies in globular clusters. The top matching words (see Figure~\ref{fig:example_papers}) are highly relevant to a topic that discusses anomalous abundances that might be caused by different pollution system to varying degrees due to their yields. The references are more similar compared to the first paper suggesting that this paper is more focused. The same is true for the citations that also come from more similar papers when compared to the citations in the first paper. The comparison of all median similarity numbers, however, shows a similar pattern when compared to the first paper including the dissimilarity of the ADS suggestion algorithm (see Figure~\ref{fig:example_papers}).

\subsection{Example: ALMA observations of a disk}

The last test paper is titled ``Unveiling the gas-and-dust disk structure in HD 163296 using ALMA observations'' \citep{2013A&A...557A.133D}. This paper describes observations of structure around young massive stars. The important words (see Figure~\ref{fig:example_papers}) again seem highly relevant. Similar to the second paper the citations and references have high similarity numbers that might suggest a narrower focus of this paper. In contrast to the other two example papers in this case the ADS suggestion algorithm also produced a high similarity index. Only three of the papers from the suggestion algorithm could be found the \arxiv corpus (compared to the usual six) which might explain this anomaly. 
 
\section{Discussion}
\label{sec:discussion}
This test of the use of natural language processing and machine learning tools \gls{nltk}, \gls{scikit-learn} has already shown that even simple techniques result in knowledge discovery.  However, there are a number of improvements that can aid in the knowledge discovery part.

Specifically, in the language processing step, there are several steps that might be improved in future versions. We remove all words that are not in the English dictionary during our initial run. This already poses some problems in the lemmatization process as the name ``Roche'' (as in Roche Lobe) is not recognized and is thus removed. This suggests that there is a need to build a domain-specific lemmatizer \citep[such as the \textsc{BioLemmatizer} for biology][]{liu2012biolemmatizer}. In our current approach, we also only consider single words (so-called unigrams) but terms like ``white dwarf'' (bigram) suggest that future iterations of this algorithm might find more relevant results if we treat such bigrams separately. Abbreviations are also commonly used in papers and are most often defined at the beginning of most papers. Thus the expanded word enters the word count only once. However, this leads to a misinterpretation of the true importance of the word as all other mentions are discarded. 

The next information carrier that is removed are object names which might link papers that are of the same object. However, using this technique already values a certain type of knowledge above another (any study on the object is valued higher than similar studies on other objects for a given paper). This is especially true in our metric as object names will have a very low document frequency (being only mentioned in few papers) and thus will attain very high values in a TFiDF comparison which might not lead to the desired result.  

\section{Conclusion}
\label{sec:conclusion}
We present a new technique for knowledge discovery by using a text similarity approach to find similar papers to a given reference paper. This technique performs robustly and finds relevant papers that are not discovered via either citations, references or suggestions from ADS. This metric also seems to be a useful tool when studying if a paper is relevant to a broader field or addresses some detail in a narrower focus. Similar attempts in other fields \citep[e.g. neuroscience;][]{2016PLoSO..1158423A} also suggest that this can be used to provide a powerful method to disseminate papers.

Currently, this allows an additional method to discover knowledge, especially when entering a field that one is unfamiliar with (e.g. using this technique for reviews). Our recommendation method might be further improved by linking our algorithm with citation information and using an algorithm like \textsc{PageRank} \citep[popularized by Google;][]{page1999pagerank}. This will value highly cited papers more than lower cited ones. While this technique will help in knowledge discovery by finding relevant papers, our future goal is to identify key measurements and statements in each paper. This will allow a scientist to quickly sift through the vast amount of knowledge and identify the relevant paper by the searched quantities (e.g. the most current mass of the proton) before reading the entire paper and critically evaluating the methodology and statistics used. 

Such a machinery would in the first instance help scientists to discover sought-after knowledge (regardless of bias towards certain authors, etc.) but might also allow for additional services. One of these might be very simple ``fact checking'' mechanisms that will aid researchers when compiling a paper by providing the most up-to-date quantities and flagging mistakes (similar to a grammar/spelling checker). 

Such a machinery has uses far beyond astronomy and astrophysics. However, among the many academic fields, astronomy exposes the vast majority of papers and data in machine-readable formats (priv. comm. Christine L. Borgman). This suggests that this field is a good start for the development of such a machinery.

\section{Acknowledgements}

W.~E.~Kerzendorf was supported by an ESO Fellowship and the Excellence Cluster Universe, Technische Universit\"at M\"unchen, Boltzmannstrasse 2, D-85748 Garching, Germany. Especially, we would like to thank the detailed discussions, encouragement, and suggestions from Felix Stoehr and Jason Spyromillio. The support from the library team (Uta Grothkopf , Dominic Bordelon \& Silvia Measkins) was invaluable to get an insight into the field of knowledge discovery. Christine Borgman and Bernie Randles (at UCLA) gave suggestion from an Information Sciences point of view and the visit would not have been possible if not for the generousity of the UCLA Galactic Center Group (especially Tuan Do and Andrea Ghez). We would also like to thank Bruno Leibundgut, Kathatrina Immer, Ivan Cabrera-Ziri for testing the algorithm on some well-known papers. Finally, we would like to thank Hinrich Sch\"{u}tze for useful discussion of tools and techniques in the NLP field.


\bibliographystyle{aasjournal}
\bibliography{wekerzendorf}

\begin{thebibliography}{}
\expandafter\ifx\csname natexlab\endcsname\relax\def\natexlab#1{#1}\fi
\providecommand{\url}[1]{\href{#1}{#1}}
\providecommand{\dodoi}[1]{doi:~\href{http://doi.org/#1}{\nolinkurl{#1}}}
\providecommand{\doeprint}[1]{\href{http://ascl.net/#1}{\nolinkurl{http://ascl.net/#1}}}
\providecommand{\doarXiv}[1]{\href{https://arxiv.org/abs/#1}{\nolinkurl{https://arxiv.org/abs/#1}}}

\bibitem[{Aboukhalil(2014)}]{10.15200/winn.141832.26907}
Aboukhalil, R. 2014, The Winnower, \dodoi{10.15200/winn.141832.26907}

\bibitem[{{Achakulvisut} {et~al.}(2016){Achakulvisut}, {Acuna}, {Ruangrong}, \&
  {Kording}}]{2016PLoSO..1158423A}
{Achakulvisut}, T., {Acuna}, D.~E., {Ruangrong}, T., \& {Kording}, K. 2016,
  PLoS ONE, 11, e0158423, \dodoi{10.1371/journal.pone.0158423}

\bibitem[{{Bastian} {et~al.}(2015){Bastian}, {Cabrera-Ziri}, \&
  {Salaris}}]{2015MNRAS.449.3333B}
{Bastian}, N., {Cabrera-Ziri}, I., \& {Salaris}, M. 2015, \mnras, 449, 3333,
  \dodoi{10.1093/mnras/stv543}

\bibitem[{Bird {et~al.}(2009)Bird, Klein, \& Loper}]{bird2009natural}
Bird, S., Klein, E., \& Loper, E. 2009, Natural language processing with
  Python: analyzing text with the natural language toolkit (" O'Reilly Media,
  Inc.")

\bibitem[{Bornmann \& Mutz(2015)}]{bornmann2015growth}
Bornmann, L., \& Mutz, R. 2015, Journal of the Association for Information
  Science and Technology, 66, 2215

\bibitem[{{Chyla} {et~al.}(2015){Chyla}, {Accomazzi}, {Holachek}, {Grant},
  {Elliott}, {Henneken}, {Thompson}, {Kurtz}, {Murray}, \&
  {Sudilovsky}}]{2015ASPC..495..401C}
{Chyla}, R., {Accomazzi}, A., {Holachek}, A., {et~al.} 2015, in Astronomical
  Society of the Pacific Conference Series, Vol. 495, Astronomical Data
  Analysis Software an Systems XXIV (ADASS XXIV), ed. A.~R. {Taylor} \&
  E.~{Rosolowsky}, 401

\bibitem[{{de Gregorio-Monsalvo} {et~al.}(2013){de Gregorio-Monsalvo},
  {M{\'e}nard}, {Dent}, {Pinte}, {L{\'o}pez}, {Klaassen}, {Hales},
  {Cort{\'e}s}, {Rawlings}, {Tachihara}, {Testi}, {Takahashi}, {Chapillon},
  {Mathews}, {Juhasz}, {Akiyama}, {Higuchi}, {Saito}, {Nyman}, {Phillips},
  {Rod{\'o}n}, {Corder}, \& {Van Kempen}}]{2013A&A...557A.133D}
{de Gregorio-Monsalvo}, I., {M{\'e}nard}, F., {Dent}, W., {et~al.} 2013, \aap,
  557, A133, \dodoi{10.1051/0004-6361/201321603}

\bibitem[{Fellbaum(1998)}]{fellbaum1998wordnet}
Fellbaum, C. 1998, WordNet (Wiley Online Library)

\bibitem[{Goulden {et~al.}(1990)Goulden, Nation, \& Read}]{goulden1990large}
Goulden, R., Nation, P., \& Read, J. 1990, Applied linguistics, 11, 341

\bibitem[{Harris(1954)}]{harris1954distributional}
Harris, Z.~S. 1954, Word, 10, 146

\bibitem[{{Henneken} \& {Kurtz}(2010)}]{2010APS..MAR.S1244H}
{Henneken}, E., \& {Kurtz}, M. 2010, in APS March Meeting Abstracts

\bibitem[{{Kerzendorf} {et~al.}(2012){Kerzendorf}, {Schmidt}, {Laird},
  {Podsiadlowski}, \& {Bessell}}]{2012ApJ...759....7K}
{Kerzendorf}, W.~E., {Schmidt}, B.~P., {Laird}, J.~B., {Podsiadlowski}, P., \&
  {Bessell}, M.~S. 2012, \apj, 759, 7, \dodoi{10.1088/0004-637X/759/1/7}

\bibitem[{{Kurtz}(2011)}]{2011ASSP...24...23K}
{Kurtz}, M.~J. 2011, Astrophysics and Space Science Proceedings, 24, 23,
  \dodoi{10.1007/978-1-4419-8369-5_3}

\bibitem[{{Kurtz} {et~al.}(2000){Kurtz}, {Eichhorn}, {Accomazzi}, {Grant},
  {Murray}, \& {Watson}}]{2000A&AS..143...41K}
{Kurtz}, M.~J., {Eichhorn}, G., {Accomazzi}, A., {et~al.} 2000, \aaps, 143, 41,
  \dodoi{10.1051/aas:2000170}

\bibitem[{Liu {et~al.}(2012)Liu, Christiansen, Baumgartner, \&
  Verspoor}]{liu2012biolemmatizer}
Liu, H., Christiansen, T., Baumgartner, W.~A., \& Verspoor, K. 2012, Journal of
  biomedical semantics, 3, 3

\bibitem[{Luhn(1957)}]{5392697}
Luhn, H.~P. 1957, IBM Journal of Research and Development, 1, 309,
  \dodoi{10.1147/rd.14.0309}

\bibitem[{Page {et~al.}(1999)Page, Brin, Motwani, \&
  Winograd}]{page1999pagerank}
Page, L., Brin, S., Motwani, R., \& Winograd, T. 1999, The PageRank citation
  ranking: Bringing order to the web., Tech. rep., Stanford InfoLab

\bibitem[{Pedregosa {et~al.}(2011)Pedregosa, Varoquaux, Gramfort, Michel,
  Thirion, Grisel, Blondel, Prettenhofer, Weiss, Dubourg, Vanderplas, Passos,
  Cournapeau, Brucher, Perrot, \& Duchesnay}]{scikit-learn}
Pedregosa, F., Varoquaux, G., Gramfort, A., {et~al.} 2011, Journal of Machine
  Learning Research, 12, 2825

\bibitem[{{Schlegel} {et~al.}(1998){Schlegel}, {Finkbeiner}, \&
  {Davis}}]{1998ApJ...500..525S}
{Schlegel}, D.~J., {Finkbeiner}, D.~P., \& {Davis}, M. 1998, \apj, 500, 525,
  \dodoi{10.1086/305772}

\bibitem[{Simpson \& Weiner(1989)}]{simpson1989oxford}
Simpson, J., \& Weiner, E.~S. 1989, Oxford: Clarendon Press. Retrieved March,
  6, 2008

\bibitem[{Sparck~Jones(1972)}]{sparck1972statistical}
Sparck~Jones, K. 1972, Journal of documentation, 28, 11

\bibitem[{van Wesel {et~al.}(2014)van Wesel, Wyatt, \& ten Haaf}]{vanWesel2014}
van Wesel, M., Wyatt, S., \& ten Haaf, J. 2014, Scientometrics, 98, 1601,
  \dodoi{10.1007/s11192-013-1154-x}

\end{thebibliography}
\end{document}